# Annotating Video with Open Educational Resources in a Flipped Classroom Scenario

Olivier Aubert, University of Nantes
olivier.aubert@univ-nantes.fr
Joscha Jaeger, Merz Akademie
joscha.jaeger@merz-akademie.de

**Abstract**

A wealth of Open Educational Resources is now available, and beyond the first and evident problem of finding them, the issue of articulating a set of resources is arising. When using audiovisual resources, among different possibilities, annotating a video resource with additional resources linked to specific fragments can constitute one of the articulation modalities. Annotating a video is a complex task, and in a pedagogical context, intermediary activities should be proposed in order to mitigate this complexity.

In this paper, we describe a tool dedicated to supporting video annotation activities. It aims at improving learner engagement, by having students be more active when watching videos by offering a progressive annotation process, first guided by providing predefined resources, then more freely, to accompany users in the practice of annotating videos.

**Introduction**

There exists a huge number of Open Education Resources (OERs) of various types, in numerous domains. OERs offer great pedagogical content, but somehow lack in articulation. One of the ways to enhance their pedagogical potential is to provide explicit links between resources, exposing some knowledge articulation, defining a path to follow or a graph of relations to explore.

This activity is at the heart of knowledge work, but its realisation as a shareable and modifiable entity is not obvious to many people. One of the goals of current teaching should be to teach students how to explicitly link these resources between themselves, so as to add explicit relationships to the existing knowledge.

Many OERs are text-based or image-based, but an increasing number relies on audiovisual content, in part due to the technological availability of capture means and the general capture policy carried out in some institutions. While hypertext links are now commonly used and understood, hypervideo links, i.e. links from/to fragments of video documents, are less common and raise a number of specific issues, given the dynamic nature of video documents. Thus while video documents are now more and more openly accessible and available, and constitute a learning material of great pedagogical value, their potential is not always fully exploited (Hobbs, 2006). Bringing the learners to a more active attitude while watching video documents can be achieved through various pedagogical practices and tools (Bossewitch, 2011; Zahn, 2010; Schwan and Riempp, 2004). We propose here a tool that can be used to promote video-based active learning, instead of passive watching, through the use of annotations, illustrated here through a flipped classroom scenario (Strayer, 2007). In addition, collaboration features promote group activities in order to invite learners to build their own shared knowledge (Scardamalia and Bereiter, 2006).

In this article, we describe a software framework dedicated to supporting video annotation activities, which can be tested at http://ocwc.open-hypervideo.org/. It aims at 1/ improving learner engagement, by having the students be more active when watching videos; 2/ offering a progressive annotation process, first guided then more free, to accompany the users in the practice of annotating videos and building hypervideos; 3/ promoting a constructivist approach by offering collaboration and discussion features.

**Video annotations and usage**

We consider here video annotations as pieces of information linked to a specific fragment of a video. They can be of different natures: textual comments, images, other audio or video resources, or more generally an URL to any resource. These contents are anchored to specific fragments of the original video document, specifying begin and end timecodes (which may be identical when the user only wants to annotate a specific point in the video instead of a durative fragment).

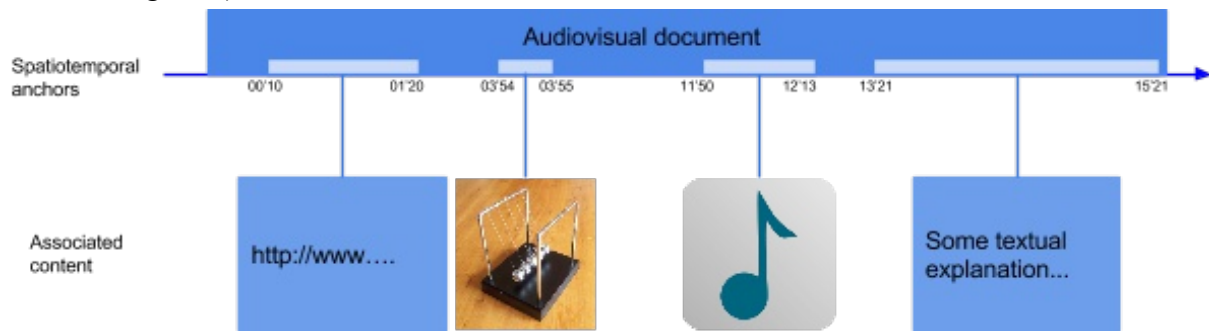

Depending on the usage context, annotations may also be associated with various metadata, such as information about their author or creation date, or some way of categorizing them.
Annotations constitute essential metadata to enable audiovisual documents handling beyond basic playing. They are required to offer indexing capabilities, either through basic chaptering of audiovisual documents or full-text search through time-aligned transcriptions for instance. They also can be used as a navigational aid for interactive video usage. Hereafter, they also can form the basis for the production of new kinds of documents, combining audiovisual documents and its metadata, which can qualify as hypervideos (Aubert and Prié, 2005).
Video annotation production and usage shares some common concerns with text or image annotation: all require anchoring schemes, raise issues around the combined evolution of both original data and its metadata, and introduce new licensing issues (which may target the document, its metadata or the combination of both). Moreover, the dynamic and temporal nature of audiovisual documents brings additional concerns, especially in annotation edition and visualisation. The complexity of generic annotation processes can thus require a fair expertise and involvement. That is why it is essential to propose more specific annotation tools, more fitted to specific tasks, in order to simplify their usage. We propose here such a tool, targeting a pedagogical context where pre-defined resources are provided and users have to assimilate them and associate them to appropriate fragments of a video.

**Usage scenario**

We will illustrate our proposal through an example usage scenario. In a science history course, a teacher records a lecture in order to use it in a flipped classroom setup: his students will have to watch the lecture recording by themselves, and they will later have a interactive session together where the teacher can answer questions.
To support this flipped classroom setup, before publishing his lecture the teacher provides additional resources such as images, texts or web resources that he publishes along the lecture recording. The learner tasks are twofold: first, they have to link the provided resources to appropriate moments of the lecture, as an exercise, through a simple interface, presented in figure 1. Second, they can annotate the recording with free-text annotations, identifying and expliciting the parts of the recording that they would like to talk about in class. The first step of the process encourages learners to be more active when watching a video, by taking into account other existing, predefined resources and linking them to appropriate parts of the lecture. Moreover, in order to properly achieve the task, they have to examine every associated resource, and memorize part of their information. The second step of the process, free-text annotation of specific lecture fragments, involves explicitation of difficult parts or of emerging ideas to be discussed in the classroom, and is useful for the teacher as a feedback, but also for the learners as a way to express concerns or ideas. The outcome of this phase then consists, for each student, in an annotated video, where pedagogical resources are synchronized with the original recording.

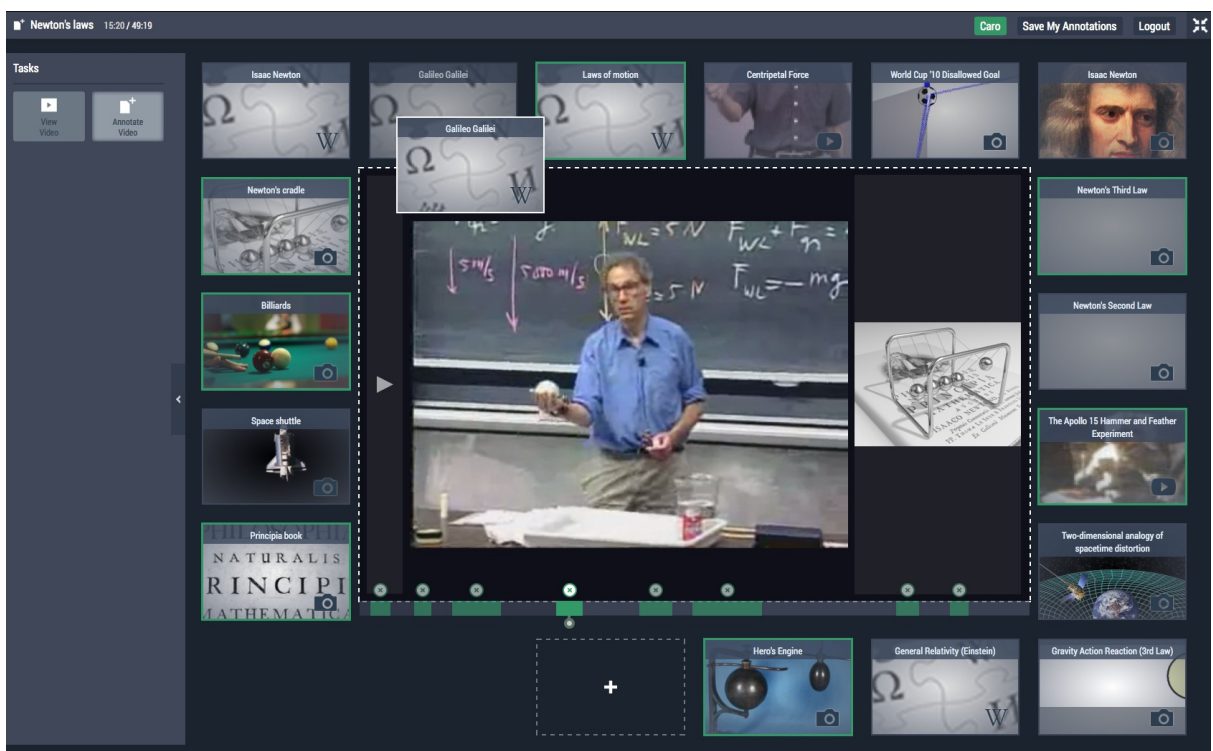

*Figure 1: resource linking interface - predefined resources have to be associated by the learners at appropriate moments of the reference video document.*

A collaboration activity can then be carried out, built up upon the produced material: the class is divided into smaller learner groups, where each member confronts his annotations (linked resources and comments) with the annotations of the other members, thanks to a simultaneous display of annotations, as shown in figure 2. Learners are then encouraged to debate on the correct timing for resources or links, in order to achieve some consensus that can be presented to the rest of the class. The integration of a revision history system could then bring a track of the different modifications, in order to possibly have the group reflect on its own group activity.

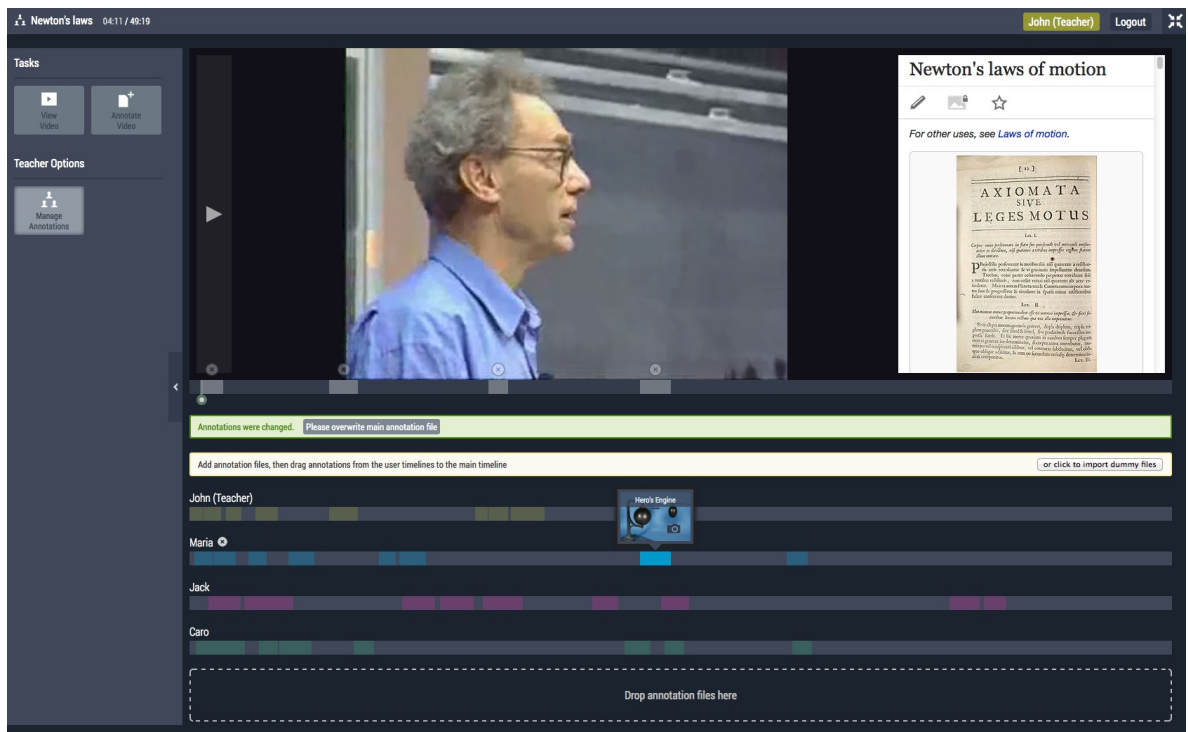

*Figure 2: collaborative comparison interface - the annotations produced by the different learners or groups are displayed simultaneously, in order to foster debate and discussion. This process leads to a consolidation of annotated resources in a main timeline, generating a joint group work that can be presented and shared with other learners.*

**Developments and future work**

The proposed tool is functional but still in development. It is built using the Open Hypervideo framework (described and available at http://www.open-hypervideo.org/) and will be published under an open-source license, making it freely reusable and modifiable, in order to be consistent with the OER idea. The architecture of the tool adopts a lightweight, componentized approach, so that it can easily be integrated into different platforms. To foster learner engagement and improve digital literacy, the tool itself is based on open web technologies and standards, enabling learners as well as teachers to use, study, remix and share the software itself as needed, as part of the outcome. Dependencies concerning the technical environment have been limited to the smallest extent possible. In this self-contained surrounding, the boundaries between creation and creational process are meant to shift, extending the definition of work to the authoring environment.

The interface provides different view modes, in which a set of options is made available based on user tasks and roles. By default, users can view the video and browse existing time-based annotations. When learners log in, the Annotate Video mode shows an overview of existing resources and facilitates adding, editing and removing annotations through a simple drag-and-drop interface. The outcome of this task is a personalised annotation set for the current video sequence, which can be exported and shared with the teacher. An additional view mode (Manage Annotations) enables importing, comparing and managing multiple annotation sets. This view mode is used to collaboratively select and merge annotations from all learner sets into one main timeline.

The view modes described above do not interfere with the video playback or time-based events. They rather provide their respective options on-the-fly, while the video continues to play. The video area is thereby always visible, scaling to provide the space needed for other interface elements.

On the technical side, the tool already supports additional means of including time-based resources, such as spatio-temporal overlays on top of the video as well as collections of linked video sequences. Designing appropriate editing interfaces for these components will be an important part of future developments. Our final goal is to expand the editing means from an annotation framework down to the filmic material. Following the principles described above, teachers and learners shall be capable of editing and remixing the underlying video sequences from within the tool. In the end, this concept leads to an agglomeration of open media fragments, assembled by programming code and thus in a constantly unfinished state. For example, a film sequence would consist of several fragments of video files, customizable overlays like text inserts or time-based effects, as well as spatio-temporal links to external media fragments. What emerges is a time-based, extendable and interactive knowledge network that seamlessly integrates with the World-Wide-Web as technical and, more importantly, as social environment.

**Discussion**

This project focuses on interface issues. However, finding appropriate open educational resources is quite time-consuming and would greatly benefit from some support from other research projects such as the LinkedUp project (Dietze et al, 2013), which aims at developing tools and methods for the exploitation of public, open data available on the Web, in particular by educational institutions. A semi-automatic approach based on semantic metadata about existing OERs could be leveraged to facilitate the finding of resources by teachers and students. To this end, our resource creation and annotation interface should be augmented to add components that assist authors in finding appropriate resources through recommendations based on semantic metadata. By working with semantic concepts and linked open data, authors will also be able to discover previously unknown relationships during the process of gathering adequate resources. Evaluating possible interfaces and interaction methods for these tasks will be part of future developments.

Video annotation is not an end in itself: it is one of the tools that is now becoming available on the pedagogical palette. Beyond video documents, annotation of any kind of document is an integral part of any scholarly work, and tools should be provided to accompany this practice on a variety of medias. Moreover, beyond the availability of such tools, appropriate integration in pedagogical scenarios and practices must be experimented and studied (Sankey and Hunt, 2013).

**Conclusion**

We propose a new open-source tool aiming at the development of new pedagogical practices, taking advantage of existing audiovisual OERs. In order to make learners more familiar with the practice of video annotation, a guided approach is proposed in a first phase, based on predefined resources proposed by the tutor. Then, learners can engage in a more open and flexible way with the video document by producing their own annotations, and carry out a collaborative activity around them. After the design and implementation phase, basic initial tests are encouraging and we are looking for more experimentation opportunities to validate our initial ideas.

References

Aubert, O. & Prié, Y. (2005) Advene: Active Reading through Hypervideos. *ACM Conference on Hypertext and Hypermedia 05*. pp. 235-244.

Bossewitch, J. & Preston, M. D. (2011) Teaching and Learning with Video Annotations. in *Learning Through Digital Media: Experiments in Technology and Pedagogy*, Chap. 19.

Dietze, S., Herder, E., Drachsler, H., d'Aquin, M., Van der Waal, S., and Parodi, E. LinkedUp – Linking Web Data for Education Project, Project Networking Track at *ESWC 2013 – 10th Extended Semantic Web Conference*, Montpellier, France, May (2013).

Hobbs, R. (2006). Non-optimal uses of video in the classroom. *Learning, Media and Technology*, 31, 35-50.

Sankey, Michael D. and Hunt, Lynne (2013) Using technology to enable flipped classrooms whilst sustaining sound pedagogy. In *30th Australasian Society for Computers in Learning in Tertiary Education Conference (ASCILITE 2013): Electric Dreams*, 1-4 Dec 2013, Sydney, Australia.

Scardamalia, M., and Bereiter, C. (2006). Knowledge Building: Theory, Pedagogy, and Technology. In *The Cambridge Handbook of the Learning Sciences* (pp. 97-118). New York.

Schwan, S., and Riempp, R. (2004). The cognitive benefits of interactive videos: Learning to tie nautical knots. In *Learning and instruction*, 14 (3), 293-305.

Strayer, J. (2007). The effects of the classroom flip on the learning environment: a comparison of learning activity in a traditional classroom and a flip classroom that used an intelligent tutoring system. (Electronic Thesis or Dissertation). Retrieved from https://etd.ohiolink.edu/.

Zahn C., Krauskopf K., Hesse F., and Pea R.. 2010. Digital video tools in the classroom: empirical studies on constructivist learning with audio-visual media in the domain of history. In *Proceedings of the 9th International Conference of the Learning Sciences - Volume 1* (ICLS '10), , Vol. 1. International Society of the Learning Sciences 620-627.

Acknowledgement

This work was partly funded by a French government support granted to the COMIN Labs excellence laboratory and managed by the National Research Agency in the "Investing for the Futures" program ANR-JO-LABX-07-0J.

License and Citation